# Single-window integrated development environment

Ivan Ruchkin
Computer Systems Lab
Moscow State University, CS department
Moscow, Russia
ruchkin.ivan@gmail.com

Vladimir Prus
Computer Systems Lab
Moscow State University, CS department
Moscow, Russia
vladimir.prus@gmail.com

*Abstract* — **This paper addresses the problem of IDE interface complexity by introducing single-window graphical user interface. This approach lies in removing additional child windows from IDE, thus allowing a user to keep only text editor window open. We describe an abstract model of IDE GUI that is based on most popular modern integrated environments and has generalized user interface parts. Then this abstract model is reorganized into single windowed interface model: access to common IDE functions is provided from the code editing window while utility windows are removed without loss of IDE functionality. After that the implementation of single-window GUI on KDevelop 4 is described. And finally tool views and usability of several well-known IDEs are surveyed.**

## I. INTRODUCTION

There is a wide variety of tools that software engineers use to write the code – from simple text editors, such as Notepad or Kate (which offer only basic text highlighting) to elaborate integrated development environments (IDEs) like Eclipse or Microsoft Visual Studio. IDEs bring together different tools improving convenience, and also providing features not possible with individual tools. For example, reliable code-completion is only possible when the editor, the compiler, and the build system work closely together.

But despite those attractive features many developers find IDE hard to use and stick with simple editors [1]. Reasons for that are different and large usability study [2] is required to completely and accurately determine them. However, anecdotic evidence shows that many users complain about the IDE tool views – which are auxiliary windows typically docked around the editor area, also called utility windows [3][4]. One problem is that a dozen of available tool views just confuse a new user. But a deeper problem is that the tool views are actually required in a number of common workflows. And if a given tool view is repeatedly required, a user is faced with unpleasant choice – either to keep the window always visible, taking space from the source editor, or to constantly open and close it. One way to address this usability issue with tool views is to remove them altogether. Of course, we propose this only as research experiment: remove all tool views, design alternative mechanism to support common workflows and then perform a usability study. These steps allow us to determine a set of tool views that users cannot live without and have to be put back. But we also hope that some utility windows can be eliminated, resulting in an overall usability improvement. One apparent example of removable tool view is build results view which shows errors and warnings. It only makes sense to process error messages from the first to the last one (as later error can be induced by the earlier) and therefore there's no necessity of showing a list of errors – we can immediately display the first error inside the text editor.

So we come to the idea of IDE with the single window hosting a text editor. Researching such approach requires:
- designing conceptual single-window IDE interface,
- implementing it in certain IDE,
- usability testing [2] of implemented GUI and comparing results to existing popular IDEs.

This paper is a work-in-progress report that covers only first two steps. The exact plan of this paper:
- Create a model of IDE tool views by observing them in the most popular development environments [1][5]. This model includes a set of abstract tool views with description of their structure and usage.
- Design a conceptual single-window IDE GUI by removing the utility windows from IDE tool view model.
- Implement the single-window interface in KDevelop integrated development environment.
- Survey existing IDE and find out whether they can be used without tool views.

It is difficult to move functionality of all tool views to text editor window. Trying to do this can cause usability problems [6]. To solve this issue we introduce new widgets and mechanisms to display information inside and near text editor. These widgets are described while designing a conceptual single-window interface.

In the following section the IDE tool view model is proposed.

## II. MODEL OF IDE TOOL VIEWS

In this section we build a model of IDE tool views. This model describes a set of abstract tool views and their functions. Each abstract tool view is generalization of similar real tool views from existing IDEs. We build such model to be able to construct a generally useful set of interface improvements, as opposed to fixing problems in a single arbitrary selected IDE.

Development process greatly depends on the programming language. To limit the scope of research we only consider development in a compiled object-oriented language such as C++, Java, or C#. According to [1] [7], the most popular IDEs for these languages are Visual Studio, NetBeans, Eclipse, KDevelop, and Code::Blocks.

First, we need to list the common parts of an IDE interface to show the context in which tool views are used. GUI of a contemporary IDE consists of:

- Main menu, typically with a vast set of commands. It isn't easy for a new developer to discover all of operations in the main menu, that fact can cause usability issues [8].
- Customizable toolbar with command buttons. It facilitates accessing and learning IDE functions, but requires a careful selection of command to expose b default.
- Text editor window. Obviously that's where program code is being edited. This area contains source code and is usually tabbed. Unfortunately, in common IDEs developer is drawn away from the text editor window because a lot of functions and information are spread across other UI elements, mostly utility windows.
- Tool views. These are additional utility child windows inside IDE GUI that contain special instruments: errors view, call graph, project view and many others. Typically views can be docked on any side of text editor window, resized and made auto hiding – popping up when mouse cursor hovers over their header and closing when it leaves. Tool views occupy considerable amount of screen space and compete for it with text editor.
- Status bar. It is a horizontal one text line-high widget in the bottom of IDE GUI that traditionally contains information on current operation progress and rarely some static information about current state of development.
- Context menus in text editor window. That's a powerful tool to provide various operations: a user accesses it quicker than the main menu and learns faster.

Due to work-in-progress paper size limit, we omit the full study of several IDEs and their tool views. The result of this study is a set of abstract tool views, collected from examined IDEs. The key abstraction points were tool view structure and functionality.

We should remark that some tool views are actually documents, displayed outside editor. A good example of document-window is "Output View" in Visual Studio. Such views should be promoted to regular documents and shown in one of text editor tabs. We take only Variables view as an example of such document-like windows, though there are more of them in existing IDEs.

Here we name only the most significant abstract tool views, which compose the tool view model:

- Project view. It is a tree view of categorized project documents. This window is used to navigate through project and interact with documents.
- Files view. It is a tree view of files in file system. This view allows a user to observe the files structure and operate over them.
- Code objects view. It is a tree view of code objects: classes, global functions, global variables and macros. Exact appearance and naming can vary depending on IDE, but all code objects views display class hierarchy as well as class methods and attributes.
- Build results view. Shows a list of build errors and warnings if there were any. Allows a user to navigate to any issue by clicking on it.
- Tasks view. Contains a list of tasks and lets a user navigate from them to text; in most cases, task is a comment from code, marked with a special word, for example "TODO", "FIXME", or "HACK".
- Breakpoint view. Shows a list of all breakpoints, allows a user to switch them on/off, edit their condition and navigate to certain breakpoint.
- Threads view. This window contains a list of debugged application threads. Clicking on a thread results in navigating to its current execution point.
- Call stack view. This view shows the stack of subroutine calls for every program thread while debugging an application. By clicking on any function call a user can open the definition of it in code editor.
- Variables view. This utility window shows a list of variables and expressions monitored while debugging and their current values.

So we have described a model of IDE tool views. This model is sufficient to introduce the single-window design.

III. CONCEPTUAL SINGLE-WINDOW DESIGN

This section shows the process and reasoning of creating a single-window IDE design based on tool view model from the previous section.

A. *Single-window design: idea*

The idea of this paper is creating a usable graphical interface design for IDE without tool views. Our goal is to reduce the number of open tool views while working with IDE and to carry over the tool views functionality into the text editor.

It is very difficult to meet this goal with the only text editor window. Thus we should introduce several new GUI elements for our single-window design. Those elements address tool view usability problems by (i) taking, when inactive, much less screen space than a tool view (or no space at all) and (ii) appearing when necessary, without explicit user interaction. These visual elements are covered in the next subsection.

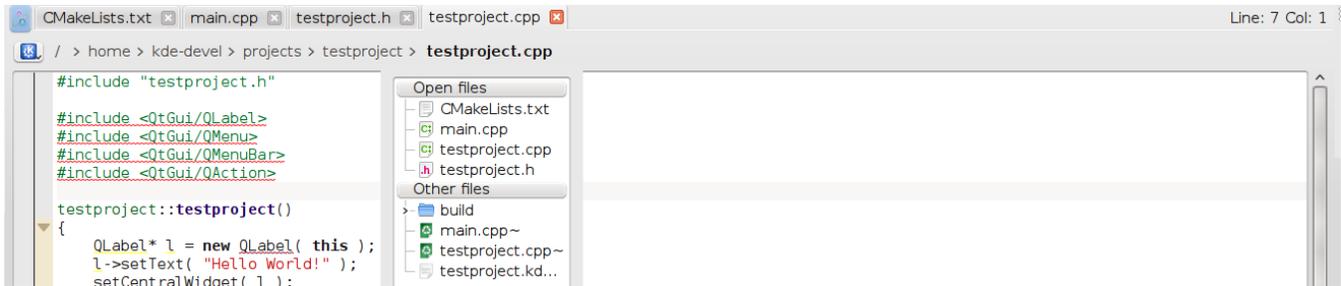

Picture 1. Breadcrumbs navigation bar.

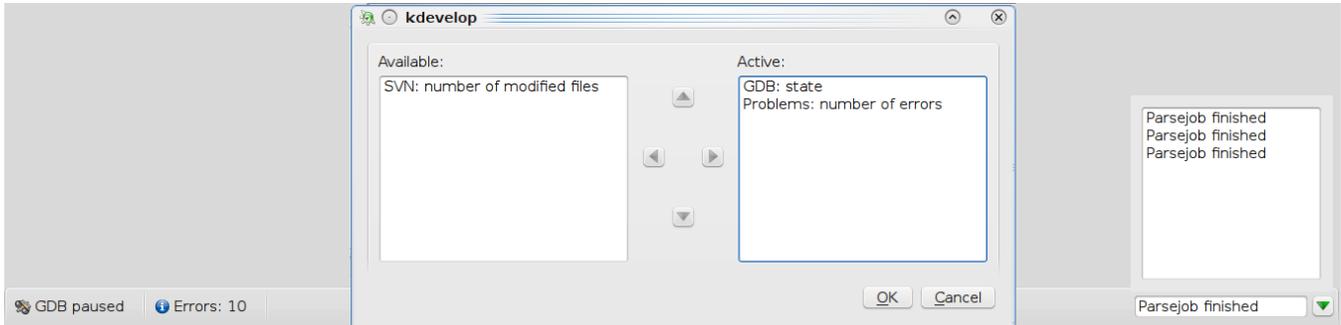

Picture 2. Enhanced status bar.

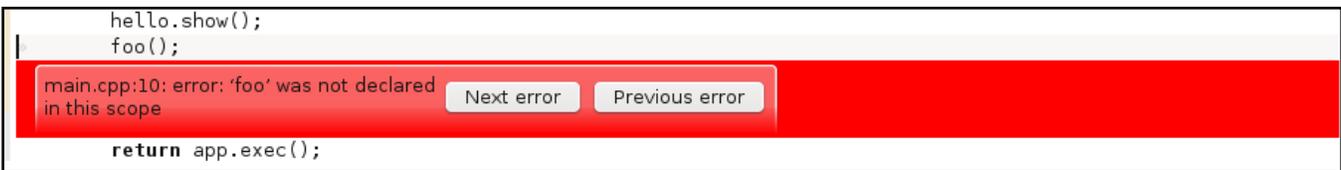

Picture 3. Inline error message.

*B. Single-window design: additional widgets*

We add widgets into single-window interface or considerably enhance the functionality of existing widgets and use them to achieve the goal from the previous subsection. Each of following paragraphs is dedicated to an additional widget.

**Breadcrumbs navigation bar**. It is a widget that represents a path in a certain tree to a currently selected object. The most common example is a file system breadcrumbs bar that shows a path to current file or directory. The bar is separated into blocks, each one representing a node in the path. Each block can be decorated with an icon or color and, on mouse click, displays a context menu with the list of elements on the same level of hierarchy. The context menu is used to change the currently selected object or to operate over other objects. The benefit of breadcrumbs lies in the principle of locality [6] [8]: a user tends to work with objects closer to his current work context. For our purposes, we propose that a single breadcrumbs bar supports different modes – that is, different meanings of what a current item mean and what hierarchy is displayed. Mode of breadcrumbs can be conventional (navigation in file system) or advanced (navigation in frame stack or classes). We describe it in details in following subsections. Breadcrumbs navigation for file path is shown on the Picture 1.

**Enhanced status bar**. Although status bar is a standard part of IDE GUI, it has to be considerably enhanced to meet the needs of single-window design. The enhanced status bar has two parts: static and dynamic. The former part is place for small (not bigger than icon plus a couple of symbols) customizable widgets that are to display information on current IDE state (for instance, number of changed files in VCS working copy). The static area should contain information that user requires occasionally [8]. The dynamic part shows messages about events in the IDE (for example, project build finished). This part shows one message at a time and can display a dropdown list of recent events so a user can choose any of them. Both status bar areas support customizable actions for left click on their parts (static widgets and messages respectively). You can see the enhanced status with several static widgets, dynamic area and setup dialog on the picture 2.

**Text editor inserts**. We want to show information linked with certain objects (variables, functions, etc.) in program code. Along with traditional ways of presenting information in text editor window (tooltips, context menus) we are going insert widgets between lines. These inline widgets can contain some information and control elements like buttons. Inline widgets are preferable when we want to display something independent on user's action. Good example is showing build errors in text (see picture 3): after build system encounters an error it should be displayed

immediately, not after user hovers mouse over error line or invokes context menu

We come to the discussion of creating single windowed interface from the model created in section II. Now we discuss each of abstract tool views listed in the previous section. We determine their main use cases and show how these tool views can be removed from user workflows.

*C. Single-window design: project view*

Project view is probably the most common type of utility window. It contains a tree of project files. Exact tree items can vary among IDEs, but the main use case is navigation through existing documents. The project view can be replaced with breadcrumbs bar. Breadcrumbs should display the path to current document, each block is a directory, and the final block is the edited document. In popup display for each block there is a tree view for directory represented by that block. In such way we allow user to navigate through project structure.

*D. Single-window design: files view*

Files view is very similar to project view in terms of structure and use cases. To replace files view we introduce a new mode for breadcrumbs. In this mode breadcrumbs bar shows a path from the root of file system to the currently edited file. Similarly to previous subsection each block is a directory, except the final one.

*E. Single-window design: build results view*

Build results window shows a list of errors and warnings. Its main use cases are viewing errors and warnings and showing corresponding source location in the editor. We replace this view functionality by displaying build issues between the lines of text and upgrading breadcrumbs bar.

Concerning errors, it's well known that the first error should be examined first, as other errors can be induced. So, it's reasonable to show the first error as inline widget, and just highlight other errors (for example, with red underlining). The inline widget should contain the text of error message, as well as "Next" and "Previous" to cycle through errors.

Warnings can't be handled like errors: their importance doesn't depend on their order. But displaying all warnings with inline widgets would takes a lot of place and can confuse a user [9]. So, we just underline code lines with warnings and display the warning text (either as inline widget or as a tooltip) only when a user clicks on the line. Also, there should be an action "Ignore this warning forever" because if warning has been shown and wasn't fixed then developer most probably won't ever touch it.

In that way the navigation is carried over to text editor. But what about overview of errors and warnings? We want to handle this use case as well. Thus we mark the blocks of file system breadcrumbs (as well as objects in popup for blocks) with red dots, reflecting that there is an error in corresponding document or that there is a document with an error in the directory.

A user also wants to have information about the number of errors and warnings. We add "Errors number" static widget to status bar. Context menu allows user to enable/disable inline text widgets and breadcrumbs marks. Also there should be a dynamic status bar message about build end with the same functionality. So the build results view has been replaced.

*F. Single-window design: code objects view*

Code objects view is used to view program structure and go to declaration of a class or any other symbol (variable, function). As shown in the previous subsection breadcrumbs bar can provide navigation through tree structures. So we can add another mode to breadcrumbs or extend the document mode: each document in the tree contains all classes, functions, global variables and macros declared in it. Thus we can remove classes view by replacing it with breadcrumbs navigation.

*G. Single-window design: tasks view*

Tasks view contains a list of tasks, which are taken from text (from comments like "// TODO" in C/C++). We replace this view similar to build results view. First, we want to carry overview functionality, so we add marks to breadcrumbs indicating that a document has a task inside. Second, we add a several characters-wide inline element after each special word ("FIXME", "TODO", etc.) with buttons for going to next and previous task. And finally we add a static status bar element that displays a number of tasks in project and allows switching on/off breadcrumbs marks and inline widgets.

*H. Single-window design: breakpoints view*

Breakpoints view contains a list of breakpoints in the project with their attributes: file, line and condition. This utility window is used when a user wants to
- create a new breakpoint,
- view breakpoints and navigate to a random breakpoint,
- edit properties of a certain breakpoint.

We want to cover all three use cases using single-window interface. Creating new breakpoints can easily be transferred to a thin vertical line on the right of text (many IDEs already maintain such operation). Overview of breakpoints can be provided by special marks in the breadcrumbs navigation bar, while it displays path to current file, just like we did it with errors in subsection III-D.

Let's assume that properties editing mostly often occurs after the breakpoint is hit. Then we handle this use case through sending dynamic message to status bar and showing inline widget with breakpoint editable properties. Navigation and global editing of breakpoints are made available through breadcrumbs like build errors (see the previous subsection).

*I. Single-window design: thread view and call stack view*

Call stack view shows called subroutines for each of program threads. We can replace both these views by organizing following tree structure: the top-level element is a thread and its descendants are function calls in stack. With that said, we introduce a new mode for viewing call stack. Breadcrumbs bar automatically switches to that mode when debugging is paused. User can handle changing modes manually through the first block of breadcrumbs. That's it for call stack view and thread view.

*J. Single-window design: variables view*

This utility window shows a list of variables and expressions that are watched during debugging process. As stated before, this view is document-like and should be placed in text editor tabs. A user wants contents of this view during debugging, so we should split the text editor window and display watched variables and expression in parallel with debugged code.

*K. Single-window design: resulting interface*

In this part of the paper we described a single-window interface design based on IDE tool views model. We have determined only some of single-window interface functional capabilities, used to carry tool views functionality to text editor and additional widgets. Much more functions can be added to that interface to improve usability. For example, contents of context menus for breadcrumbs can be prioritized upon user's selections, but such enhancements are outside of this paper's topic.

Let's proceed to the implementation details of single-window interface.

## IV. IMPLEMENTATION DETAILS

The single-window design has been implemented on the base of KDevelop 4 IDE, part of KDE project [10]. The source code of all components of KDevelop is open, which facilitated implementation of desired modifications. Implemented GUI parts can be seen on pictures 1 – 3 above.

*A. KDevelop 4 GUI*

KDevelop GUI consists of interface parts that have been described in section II. The main window consists of dockable tool views and tabbed text editor view. KDevelop tool views include "File System" view, "Project" view, "Errors" view, "Breakpoints" view. Status bar is almost empty, containing only the progress bar for showing pending operations progress, so this space can be used to implement enhanced status bar. To plug text editor inside the main window KDevelop uses KParts [11], thus allowing a user to edit text with Kate editor [12]. So, single-window interface can be implemented in KDevelop.

*B. Implementation*

To create single-window GUI in KDevelop the following components have been implemented:
- breadcrumbs navigation bar,
- showing of errors and warnings between lines,
- extended status bar.

Breadcrumbs navigation bar is placed above the code editing window and shows the path to currently edited document. When user clicks any node of the bar, a context menu with file system for that node appears. To speed up access to frequently opened documents the list of currently opened files appears above other files.

Errors and warnings are shown in following way: only the first one appears after the line it refers to. The decision to show only one line has been made because programmer usually tries to fix the first errors as later ones can be caused by the first one. Along with the error information two navigation buttons are shown: "Next" and "Previous". These buttons can take user to neighboring errors.

Enhanced status bar meets the description from section III: it has dynamic and static parts. The static part accepts small widgets for changed files in working copy, number of tasks and number of background parser errors. Presence and order of these widgets can be customized by user. The dynamic part shows messages from removed tool views: that build is started and finished, messages from debugger, background parsing results. Clicking on a message allows user to open the tool view that sent the message.

Here follows a survey of the most popular IDEs.

## V. EXISTING IDES

This section examines the sets of tool views and their usability in several popular IDEs [1][5][7]. Along way we note whether any of these IDEs can be used without tool views.

*A. Visual Studio*

Microsoft Visual Studio [13] is one of the most popular commercial IDEs for developing in the C++ and C# programming languages. Visual Studio has several often used tool views described below.
- "Solution Explorer" – a view displaying the tree of projects and files in current solution. A user operates with this view to create new files, open files (unless they are present in text editor's tabs), and observe the structure of projects.
- "Class View" is a two-part window with a list of classes in the first part and contents of classes (attributes and methods) in the second part. This window is useful for looking through class structure and navigating to their declaration or their methods implementation in code.
- "Output View" – a view containing the results of code build with errors and warnings among them or program execution results. This window is vital for building process because a user views errors and navigates to them with this window.
- "Code Definition Window" – a utility window, browsing the definition for the currently selected object or function. It is used to quickly look at and probably edit the definition of class or function.

Tool views in Visual Studio behave like described in section II and can be put in auto-hide mode, in which they expand only mouse cursor moves to their header and collapse to the screen border when cursor leaves them. That decreases the time user spends with tool views, but user can't remove them permanently: too much information is concentrated in them. For example, user can't open a new file without "Solution Explorer" and can't navigate through errors without the "Output" view. So there's no way to use Visual Studio without tool views.

*B. Eclipse*

Eclipse [14] is an open-source cross-platform IDE, mainly used for C++ and Java development. Tool views organization in Eclipse is similar to the one in Visual Studio but Eclipse auto-hide mode is less usable than in Visual Studio: it requires an extra click to open a hidden tool view. Also, a user can only toggle auto-hide mode of a certain tool view, not a whole dock area, like in Visual Studio.

Eclipse tool views resemble Visual Studio tool views. "Navigator", "Outline" and "Make targets" windows provide the navigation service through project files, identifiers, and make targets correspondingly. "Problems" and "Console" views are similar to Visual Studio "Output" view: they allow a user to look at errors and program output and go to errors in code. Thus, tool views in Eclipse play an important role in user's workflow and cannot be removed without replacing their functions with some other interface elements.

*C. Code::Blocks*

Code::Blocks [15] is an open source cross platform IDE, which is designed for convenient usage with different C++ libraries: GTK+, Qt4, OpenGL, FLTK, wxWidgets, Lightfeather. The user interface of Code::Blocks is less complex (fewer tool views and less their customizability) than one of Visual Studio and Eclipse, but that's just because of lower number of integrated features. Code::Blocks GUI has several tool views necessary for normal work:

- "Projects" – a window for tree navigation through opened projects. This is counterpart of Visual Studio "Solution Explorer" and Eclipse "Navigator" window.
- "Symbols" – a tree navigation window through functions, classes and global variables. It is similar to Visual Studio "Class View".
- "Logs" is a tabbed set of several windows: "Search results" (the list of found items), "Build log" (plaintext output of build tool), "Build messages" (a clickable list of errors and warnings), and some others.

Code::Blocks has no auto-hide mode for tool views, and this makes its interface even less suitable for reducing user's interaction with tool views.

*D. NetBeans*

NetBeans [16] is an open source cross platform IDE, used massively for development on Java platform, but also can be used for some other languages including C++.

Tool views in NetBeans are organized just like in Visual Studio. The most used tool views are:

- "Projects" and "Files" windows – tree views for navigating through projects and file system.
- "Classes" – a window for viewing classes and functions, like "Class View" in Visual Studio and "Symbols" in Code::Blocks.
- "Build" – a window with build results log, allowing a user to move to any error or warning.
- "Navigator" window – a dynamic version of classes window, shows at which place in class and function tree a user's cursor is situated. This window is useful while browsing through highly nested code.

Tool views in NetBeans can be put in auto-hide mode, called "Minimized" in the NetBeans interface. Unfortunately, a user can change the mode of only one window at a time. Tool views contain information and operations, which are essential for development process (for example, opening a new file through "Files" window), so IDE can't be used without constant interaction with utility windows. So we can conclude that NetBeans IDE envisages use of tool views in common developer workflows.

VI. CONCLUSION AND FUTURE WORK

This paper has proposed the single-window user interface for IDE to solve the usability problem of tool views. The first result is a model of IDE tool views. The second result of this paper is the conceptual design of single-window IDE interface. Key GUI elements of this interface are:
- breadcrumbs navigation bar,
- enhanced status bar,
- text editor window with inline message display functionality.

The third result of this work is the partial implementation of this single-window design in KDevelop IDE.

Future work involves usability testing [2][6] of single-window interface to find whether usability problems are solved or at least reduced compared to traditional IDE GUI.

REFERENCES

[1]  Developpez LLC, "Les meilleurs environnements de developpement" [HTML] (http://general.developpez.com/edi/)
[2]  J. Nielsen, "Usability Engineering", Academic Press, 1993, pp. 23 – 37, 165 – 227.
[3]  M. Szymczyk, "Reducing XCode's Window Clutter", 2007, [HTML] (http://meandmarkpublishing.blogspot.com/2007/06/reducing-xcodes-window-clutter.html)
[4]  Website article: M. Stephens, "10 Things NetBeans Must Do to Survive", 2003 [HTML] (http://www.softwarereality.com/soapbox/netbeans.jsp)
[5]  M. Caron, "Survey on Usability of Integrated Development Environment" [HTML] (http://docs.google.com/present/view?id=addqfjnjc3d6_108gj3w67c3)
[6]  Steve Krug, "Don't Make Me Think A Common Sense Approach to Web Usability", Indianapolis: New Riders, 2000, pp. 10 – 20, 50 – 96, 138 – 174.
[7]  Janel Garvin, "Software Development Platforms - 2009 Rankings", Evans Data Corporation, 2009, [HTML] (http://www.evansdata.com/reports/viewRelease_download.php?reportID=19)
[8]  Morgan Kauffman, "GUI Bloopers 2.0 Common User Interface Design Don'ts and Dos.", Morgan Kauffman Publishers, 2007, pp. 7 – 51.
[9]  Donald A. Norman, "The Design of Everyday Things", Doubleday, 1989, pp. 187 – 219.
[10] KDE Community, KDevelop website [HTML] (http://www.kdevelop.org/)
[11] Philippe Fremy, "KDE Technology: KParts Components" [HTML] (http://phil.freehackers.org/kde/kpart-techno/kpart-techno.html)
[12] Kate webstite [HTML] (http://kate-editor.org/)
[13] Microsoft, MSDN, Microsoft Visual Studio [HTML] (http://msdn.microsoft.com/ru-ru/vstudio/default.aspx)
[14] Ecliplse Foundation, Eclipse website [HTML] (http://www.eclipse.org/)
[15] Code::Blocks website [HTML] (http://www.codeblocks.org/)
[16] NetBeans website [HTML] (http://netbeans.org/)